\documentclass
[nofootinbib,superscriptaddress,aps,prd,showkeys,noshowpacs,onecolumn,10pt]{revtex4-2}%
\usepackage{graphics}
\usepackage{graphicx}
\usepackage{subcaption}
\usepackage{epsf}
\usepackage{bm}
\usepackage{amsmath,amssymb,amsfonts,mathrsfs,amsthm}
\usepackage{latexsym}
\usepackage{enumerate}
\usepackage{comment}
\usepackage[dvipsnames,svgnames,x11names,table]{xcolor}
\usepackage[colorlinks = true,
            linkcolor = Cerulean,
            urlcolor  = magenta,
            citecolor = magenta,
            anchorcolor = NavyBlue]{hyperref}
\usepackage{epstopdf}%
\usepackage{bbm}
\def\be{\begin{equation}}
\def\ee{\end{equation}}

\makeatother

\begin{document}

\title{Nontrivial Symmetries in $k$-essence Cosmology}
\author{Andr\'es Lueiza-Colip\'i}
\email{a.lueiza01@ufromail.cl}
\affiliation{Departamento de Ciencias F\'{\i}sicas, Universidad de La Frontera, Casilla
54-D, 4811186 Temuco, Chile}
\author{Nikolaos Dimakis}
\email{nikolaos.dimakis@ufrontera.cl}
\affiliation{Departamento de Ciencias F\'{\i}sicas, Universidad de La Frontera, Casilla
54-D, 4811186 Temuco, Chile}
\author{Andronikos Paliathanasis}
\email{anpaliat@phys.uoa.gr}
\affiliation{Institute of Systems Science, Durban University of Technology, Durban 4000, South Africa}
\affiliation{Centre for Space Research, North-West University, Potchefstroom 2520, South Africa}
\affiliation{Centro de Investigaci\'on, Innovaci\'on y Creaci\'on (CIIC), Universidad Cat\'olica de Temuco, Temuco, Chile}
\affiliation{Departamento de Ciencias Matem\'{a}ticas y F\'{\i}sicas,  Facultad de Ingenier\'{\i}a, Universidad Cat\'olica de Temuco, Temuco, Chile}
\affiliation{National Institute for Theoretical and Computational Sciences (NITheCS),
South Africa}

\begin{abstract}\noindent
In the context of a spatially flat FLRW
background, we perform a symmetry classification of $k$-essence models with
Lagrangian densities of the form $f_1(\phi)R+f_2(\phi,X)$. The kinetic term $X$ is introduced as an independent degree of
freedom via a Lagrange multiplier, and the lapse function is treated as a
dynamical variable. The symmetry analysis is applied to the
constrained system prior to any gauge fixing. This approach reveals
symmetries which otherwise are lost when the lapse is fixed at the level of the
action. The derived families of $k$-essence models admitting nontrivial symmetries fall to two general classes: minimally coupled and nonminimally coupled theories to gravity. We explore the correspondence between the symmetries through conformal mappings and use the corresponding Noetherian conservation laws to derive exact cosmological solutions. We find that, when the numerical value of the conserved charges is zero, the field equations reduce to an algebraic relation. We subsequently derive power-law expressions for the scale factor with exponents determined by the particular $k$-essence function. 
\end{abstract}

\maketitle

\section{Introduction}

The publication and analysis of the latest observations from the Dark Energy
Spectroscopic Instrument (DESI) \cite{des1,des2} challenge the standard $\Lambda $CDM model of cosmology regarding the late-time evolution of the
universe, suggesting a preference for a dynamical dark energy
scenario \cite{des3}. Furthermore, to address a series of fundamental cosmological problems, it is postulated that the universe underwent another
accelerated phase in the past, known as cosmic inflation \cite{inf1,inf2}. During this accelerated phase, the universe expanded so rapidly that it effectively lost any memory of its initial conditions. Nevertheless, the exact mechanism responsible for the description of the early- and late-time acceleration phases of the universe remains unknown. 

Cosmologists have proposed a plethora of models to explain
the observations, introducing new dynamical degrees of freedom associated
either with scalar fields, as in quintessence \cite{quin}, phantom \cite{pha}, and related models \cite{kk,bd1,bd2,bd3}, or with geometric modifications
of gravity, as in extended and alternative theories of gravity \cite{ff1,ff2,ff3,ff4,ff5}. In order to explore the physical properties
of these models, it is important to understand the behavior of their
solution trajectories using analytical or numerical techniques. In this
context, exact and analytic solutions play a crucial role in the study of the
dynamics of cosmological models. While numerical solutions provide the behavior
of the system for specific choices of the model parameters and initial
conditions, analytic solutions can reveal the general properties of the
underlying dynamics. Specifically, they can provide important information
regarding the initial value problem, the essential degrees of
freedom, and the asymptotic dynamics at different scales of the universe and
close to the cosmological singularity. Therefore, the derivation of exact
and analytic solutions is essential to examine the validity of a
cosmological theory and explore all its physical properties in depth.

In the framework
of General Relativity (GR), homogeneous cosmological models belonging to the family of Bianchi Class A geometries possess an equivalent minisuperspace description. In this
formulation, the gravitational field equations can be interpreted as the
equations of motion of a particle moving in a curved space, known as the
minisuperspace, under the influence of a potential term \cite{mp1,ns7a}. The
coordinates of this configuration space are given by the cosmological scale
factors together with the dynamical degrees of freedom associated with the dark
energy sector. In many cases, interesting analogies between gravitational and mechanical systems can be drawn. For instance, the Friedmann--Lema\^{\i}tre--Robertson--Walker (FLRW) spatially flat $\Lambda $CDM model can be mapped to the linear equation of a ``hyperbolic-oscillator'', while in the case of the vacuum Bianchi type I model leading to the Kasner solution \cite{kas1}, the three scale factors can be viewed as three free particles in a flat space with a vanishing total ``energy''.  

The minisuperspace description is also essential in the formulation of quantum cosmology. The resulting Wheeler-DeWitt equation \cite{deWitt,whe}, which
follows from the Hamiltonian constraint of the gravitational model, is
streamlined into a single equation analogous to the Schr\"{o}dinger
equation in Quantum Mechanics \cite{ha1,ha2,ha3}. Furthermore, the existence
of a minisuperspace enables the application of well-known mathematical techniques from
analytical mechanics in the description of the gravitational dynamics. By
adopting the minisuperspace description it was found in \cite{com1,com2,an1} that
various cosmological and gravitational models are invariant under a specific
group of transformations, which allows the field equations to be linearized
through the Eisenhart--Duval lift. In this way, distinct solutions can
be traced back to the same origin, namely the solution of the linearized
system corresponding to a free particle. The solutions of the original
dynamical system can then be obtained through an appropriate coordinate
transformation, which relates the original variables to the linearized ones. 

When exploring the transformations that leave the action form invariant, Noether's theorem can be employed to determine the
corresponding conservation laws. These are useful both for the construction of
exact and analytic solutions as also for revealing important information regarding the integrability of the gravitational model \cite{ns1}. Noether's
theorem has also been employed within the Ovsiannikov's scheme \cite{ovs}
for the classification of different gravitational models, imposing
constraints on the free parameters and functions of the corresponding
theories (see, for instance \cite{ns1,ns2,ns3,ns4,ns5,ns6,ns7,ns8,ns9,ns10,ns11} and
references therein). The Noetherian conservation laws are also important in quantum
cosmology. They allow us to identify quantum observables commuting with the Hamiltonian, which are necessary for the quantization of the underlying
classical system \cite{qn1,qn2,qn3,qn4,qn5}.  

In this study, we present a detailed classification of $k$-essence
cosmological models according to their admitted nontrivial symmetries within
the framework of Ovsiannikov's scheme. $k$-essence theories provide a
framework that can describe both dark matter and dark energy, while also
offering a possible unified description of the inflationary epoch and the
late-time accelerated expansion of the universe (see \cite%
{Bose:2008ew,Bose:2009kc}). Special classes of $k$-essence models include tachyon fields \cite{Bagla:2002yn}, as well as simple quintessence models. A
geometric construction of the $k$-essence theories follows from an extended
geometric framework of gravity, where the Levi-Civita connection is replaced
with the Schr\"{o}dinger connection \cite{Csillag:2025gnz}. Recently, in 
\cite{ka1}, the Dirac-Bergmann algorithm \cite{Dirac,Bergmann,Diracbook} was applied in the canonical
quantization of a class of $k$-essence models. Furthermore, in \cite{ka2}
the cosmological dynamics of a $k$-essence model with a nontrivial symmetry
configuration were examined, showing that the model can well describe
the early time inflationary epoch, proving an exit from inflation with the $k
$-essence field evolving into a pressureless dark matter component. For further
applications of $k$-essence models in gravitation and cosmology we refer to \cite%
{ke1,ke2,ke3,ke4,ke5,ke6,ke7,ke8,ke9,ke10} and references therein. 

The general Lagrangian of $k$-essence theory, $\mathcal{L}(R,\phi ,X)$, involves an arbitrary function of the Ricci scalar $R$, the scalar field $\phi$ and its kinetic term $%
X=-\frac{1}{2}\partial _{\mu }\phi \partial ^{\mu }\phi $. It is therefore  important to establish a selection rule to constrain this infinite-dimensional function space. Symmetries can provide such a geometric selection rule, since their generators are determined by the geometry of the
corresponding minisuperspace. Consequently, the requirement for the
existence of nontrivial symmetries imposes a self-consistent geometric
constraint on the allowed forms of the theory \cite{ns7}. 

The structure of the paper is as follows: In Section \ref{section2} we present the basic properties and definitions for the variational symmetries of the action. The $k$-essence theory is reviewed in Section \ref{section3}, where we consider
the families of models where the Lagrangian is linear to the Ricci scalar and
coupled only to the field $\phi$ and not to the kinetic term. These models form the simplest modifications to GR, which it is desirable to be recovered at some limit. For this
type of systems, we write the cosmological field equations and derive the
equivalent point-like Lagrangian for the minisuperspace description. Section %
\ref{section4} forms the main core of this work, where we present the
complete symmetry classification for the $k$-essence models. We compare our
results with previous studies and show that the treatment we follow here produces new cases, providing thus a complete treatment of the problem. The resulting models are divided into two
broad families: theories minimally and
nonminimally coupled to gravity. In\ Section \ref{section5} we demonstrate
the application of the Noetherian charges for the derivation of exact
cosmological solutions. Finally, in Section \ref{section6} we summarize our
results and draw our conclusions.

\section{Variational symmetries of the action} \label{section2}

To establish the general theoretical setting, let us start from a given Lagrangian $L=L(Q,\dot{Q},t)$, where the $Q^I$ denote the degrees of freedom of the problem. We assume that they are $d+1$ in number, with the index $I$ taking  values $I=0,...,d$. Form invariance of the respective action under a symmetry transformation is equivalent to the condition \cite{Olver,Stephani}
\begin{equation} \label{infcrit}
  \mathrm{pr}^{(1)} \xi (L) + \chi \frac{dL}{dt} = \frac{dF}{dt} , 
\end{equation}
which is known as the infinitesimal criterion of invariance. The vector
\begin{equation}
  \xi = \chi \frac{\partial}{\partial t} + \eta^I \frac{\partial}{\partial Q^I},
\end{equation}
is the generator of the symmetry transformation, while the
\begin{equation} \label{prolog}
   \mathrm{pr}^{(1)} \xi = \xi + \phi^I \frac{\partial}{\partial \dot{Q}^I} ,
\end{equation}
denotes its first prolongation; that is, its extension to the space of the derivatives $\dot{Q}^I=\frac{dQ^I}{dt}$, with the relevant components being calculated from
\begin{equation} \label{prologcoef}
   \phi^I = \frac{d}{dt} \left( \eta^I - \dot{Q}^I \chi \right) + \ddot{Q}^I \chi .
\end{equation}
The function $F$ in the symmetry condition \eqref{infcrit} is known as the gauge function, and reflects the freedom of the action to remain form invariant under a transformation up to the addition of a surface term. When $F$ is not trivial, $F\neq$const., the vectors $\xi$ satisfying \eqref{infcrit} are sometimes referred to as quasi-symmetries, with the term symmetries being reserved for the $F=$const. case.

For finite-dimensional symmetry groups, whose generators $\xi$ are solutions of \eqref{infcrit}, the corresponding conserved charges are obtained by the formula
\begin{equation}
  I = \eta^I \frac{\partial L}{\partial \dot{Q}^I} - \chi\left(\dot{Q}^I \frac{\partial L}{\partial \dot{Q}^I} - L \right) -F,
\end{equation} 
where we recognize $p_I=\frac{\partial L}{\partial \dot{Q}^I}$ as the conjugate momenta, while in the parenthesis there appears the Hamiltonian. According to Noether's first theorem, we obtain that $\frac{dI}{dt}=0$ holds on mass-shell, that is upon satisfaction of the Euler-Lagrange (E-L) equations.

Noether's second theorem on the other hand, does not refer to conserved quantities. It states that the existence of an (uncountably) infinite symmetry group implies that not all E-L equations are independent, a feature that is particularly relevant for cosmological Lagrangians.

In the cosmological minisuperspace description, a Lagrangian of a finite-dimensional system is considered valid if it reproduces correctly the result of the field equations \cite{mp1}. Such Lagrangian functions have a similar form even for distinct gravitational theories:
\begin{equation} \label{Lagcosmo}
  L = \frac{1}{2N} G_{ij} \dot{q}^i \dot{q}^j - N V(q) ,
\end{equation}
where $N$ is the lapse function of the metric and $q^i$ represents the rest of the degrees of freedom (scale factors, matter fields, etc.); here we assume $i=1,...,d$. So, in total we have $Q^I=(N,q^i)$, the $d+1$ degrees of freedom we mentioned earlier. 

All Lagrangians of the form \eqref{Lagcosmo} admit the infinite-dimensional symmetry generated by \cite{ns8}
\begin{equation} \label{Xinf}
  \xi_{\infty} = \chi(t) \frac{\partial}{\partial t} - \dot{\chi}(t) N \frac{\partial}{\partial N},
\end{equation}
with $\chi(t)$ being an arbitrary function of time. The above symmetry transformation is a time re-parametrization
\begin{equation}
  t\rightarrow \tilde{t} = f(t), \quad N \rightarrow \tilde{N}(\tilde{t}) = N(t) \frac{dt}{d\tilde{t}} = \frac{1}{\dot{f}}N,
\end{equation}
which leaves form-invariant the action of the system. This transformation is what remains when reducing to the mini-superspace description from the diffeomorphism invariance of GR, or of any gravitational theory whose action is composed by scalars. As previously stated, the existence of this symmetry, through Noether's second theorem, implies that not all of the Euler-Lagrange equations are independent. True enough, the set of the equations of motion of \eqref{Lagcosmo} consists of $d$ second order equations for the $q_i$'s and a quadratic constraint equation for the lapse $N$. The existence of this latter relation results in only $d-1$ acceleration involving equations being independent. To see this, solve the quadratic constraint equation algebraically with respect to $N$ and substitute the result into the $d$ second order equations for the $q_i$'s. The ensuing set is solvable algebraically only with respect to $d-1$ accelerations, leaving one of the $q_i$'s arbitrary. This represents the gauge fixing freedom, which allows to consider one of the degrees of freedom as an effective time variable. 

The property of parameterization invariance has some interesting implications in the symmetry structure of $L$ with respect to the existing conserved quantities. To understand the difference let us briefly consider the Lagrangian stripped from this freedom. Take for example the gauge fixed version of $L$, where we have set $N=1$,
\begin{equation}
  L_{N=1} = \frac{1}{2} G_{ij} \dot{q}^i \dot{q}^j - V(q).
\end{equation}
This Lagrangian has no constraint equation and its acceleration involving equations are all independent.
The symmetries of $L_{N=1}$ leading to conserved quantities will in general be fewer than those of the original Lagrangian \eqref{Lagcosmo}. The reason behind this rests precisely on the fact that all of the Euler-Lagrange equations of $L_{N=1}$ are independent. The Noether symmetry algorithm, when applied to $L_{N=1}$, reveals the conserved quantities for a system of $d$ (second order) independent equations, while in the case of \eqref{Lagcosmo} it reveals those admitted by $d-1$ independent equations. Consequently, the symmetry structure of \eqref{Lagcosmo} is in many cases richer than that of $L_{N=1}$ \cite{ns8}. In the phase-space description, these extra symmetries, whose existence is owed to the presence of constraints, were denoted by Kucha\v{r} as conditional symmetries \cite{Kuchar}. 

To clarify this point let us note that the process of gauge fixing performed at the level of solving the equations is truly inconsequential. However, when applied at the level of the action, and before searching for symmetries, it may lead to an over-restriction of the problem.

We shall proceed to explore the symmetries of $k$-essence theories that are linear in the Ricci scalar, using a parametrization invariant Lagrangian of the form \eqref{Lagcosmo}. As we shall demonstrate, this approach reveals novel classes of symmetries that have not been previously reported in the literature, precisely because earlier studies restricted the problem by using a gauge fixed Lagrangian like $L_{N=1}$. In the next section, we focus on the derivation of the minisuperspace Lagrangian.

\section{$k$-essence minisuperspace} \label{section3}

The most general $k$-essence theory has a Lagrangian density of the form $\mathcal{L}(R,\phi,X)$, where $R$ is the Ricci scalar, $\phi$ the scalar field and 
\begin{equation} \label{Xdef}
X=-\frac{1}{2}\partial_\mu\phi \partial^\mu \phi
\end{equation}
its kinetic term. For simplicity we employ the short-hand notation $\partial_\mu=\frac{\partial}{\partial x^\mu}$. As we previously mentioned, we restrict our study to milder modifications of GR of the form
\begin{equation} \label{actgen}
  S = \int\!\left[f_1(\phi) R + f_2(\phi,X) \right] \sqrt{-g} \, d^4x .
\end{equation}
The field equations for the above action are given by \cite{Kobayasi}:
\begin{equation} \label{feq1}
  f_1(\phi) G_{\mu\nu}-\left(\frac{1}{2} f_2(\phi,X)- \Box f_1(\phi)\right) g_{\mu\nu} - \nabla_\mu\nabla_\nu f_1(\phi)  - \frac{1}{2}  f_{2,X}(\phi,X) \partial_\mu \phi \partial_\nu\phi =0,
\end{equation}
for the metric, and 
\begin{equation} \label{feq2}
  \nabla_\mu \left[f_{2,X}(\phi,X) \nabla^\mu \phi \right] + R f_{1,\phi}(\phi) + f_{2,\phi}(\phi,X) = 0,
\end{equation}
for the scalar field. In the previous expressions, we introduce the notation $ f_{1,\phi}(\phi)= \frac{d f_1}{d\phi}$, $f_{2,X}(\phi,X)= \frac{\partial f_2}{\partial X}$ and $f_{2,\phi}(\phi,X)= \frac{\partial f_2}{\partial \phi}$.

Assuming a spatially flat Friedmann-Lemaître-Robertson-Walker (FLRW) spacetime 
\begin{equation} \label{FLRW}
  ds^2=-N(t)^2 dt^2 + a(t)^2 \left(dx^2+dy^2+dz^2\right) ,
\end{equation}
and a scalar field depending only on time, $\phi(t)$, a minisuperspace Lagrangian can be extracted from the starting action \eqref{actgen}. In the former, the validity of definition \eqref{Xdef} is enforced through the inclusion of a Lagrange multiplier $\lambda$. After removing a total derivative, and integrating out the spatial degrees of freedom, we are left with the following expression \cite{ka1}
\begin{equation}
  L_\lambda = -\frac{6}{N}\left(a^2  f_{1,\phi}(\phi)\dot{a} \dot{\phi}+a  f_{1}(\phi)\dot{a}^2 \right)+ a^3 N f_{2}(\phi,X)+ \frac{1}{2} \lambda \left(\frac{\dot{\phi}^2}{N^2}-2 X\right)
\end{equation}
To simplify the problem and eliminate the Lagrange multiplier, we calculate its value through the Euler-Lagrange equation for $X$, namely $\frac{\partial L_\lambda}{\partial X}=0$. We then substitute it back into $L_\lambda$ obtaining the final Lagrangian, which is of the form \eqref{Lagcosmo}
\begin{equation} \label{Lag}
  L= \frac{1}{2 N} \left(-12 a f_{1}\dot{a}^2 -12 a^2  f_{1,\phi}\dot{a} \dot{\phi}+ a^3 f_{2,X}\dot{\phi}^2 \right) + N a^3 \left( f_{2} - X f_{2,X} \right).
\end{equation}
The Euler-Lagrange equations for the degrees of freedom $N$, $a$, $\phi$ and $X$ are equivalent to \cite{ka1}
\begin{align} \label{con}
  & \frac{1}{N^2}\left[\frac{6  f_1 \dot{a}^2}{a^2}+\frac{6 f_{1,\phi}\dot{a} \dot{\phi} }{a} -\frac{f_{2,X}\dot{\phi}^2}{2}   \right]+ f_2-X f_{2,X} =0 \\ \nonumber
  & \frac{4 f_1 \ddot{a} }{a N^2} + \frac{2 f_{1,\phi}\ddot{\phi} }{N^2} + \frac{2 f_1 \dot{a}^2 }{N^2 a^2 } + \frac{\left(f_{2,X}+4 f_{1,\phi\phi}\right)\dot{\phi}^2 }{2 N^2} + \frac{4 f_{1,\phi}\dot{a} \dot{\phi} }{a N^2}-\frac{4 f_1 \dot{N}\dot{a} }{a N^3} -\frac{2 f_{1,\phi}\dot{N} \dot{\phi} }{N^3}\\
  & +  f_2 -X f_{2,X}  =0  \\ \nonumber
  & \frac{6 f_{1,\phi}\ddot{a}}{N^2}-\frac{a f_{2,X}\ddot{\phi} }{N^2}+\frac{6 f_{1,\phi} \dot{a}^2 }{ N^2 a} -\frac{3 f_{2,X} \dot{a} \dot{\phi} }{N^2} + \frac{a f_{2,X} \dot{N} \dot{\phi} }{N^3}-\frac{a f_{2,XX} \dot{X} \dot{\phi} }{N^2}-\frac{a f_{2,X\phi} \dot{\phi}^2 }{2 N^2}\\
  & -\frac{6 f_{1,\phi} \dot{a} \dot{N} }{N^3} +a \left(f_{2,\phi}-X f_{2,X\phi}\right) =0 
\end{align}
and
\begin{equation} \label{defX}
  X = \frac{\dot{\phi}^2}{2 N^2},
\end{equation}
with this last equation being obtained under the condition that $f_{2,XX}\neq 0$. That is, we exclude theories linear in the kinetic term $X$. We follow this assumption in order to work with a pure $k$-essence theory leaving out the rather trivial reduction to the quintessence case. 

It can be straightforwardly checked that the above E-L equations correctly reproduce the result of the field equations \eqref{feq1} and \eqref{feq2}, when the metric of \eqref{FLRW} is substituted and $\phi=\phi(t)$ is assumed. Thus, Lagrangian \eqref{Lag} is valid and correctly reproduces the dynamics of the gravitational system. Any symmetries obtained for the variational problem set by $L$ are going to automatically yield conserved charges for the gravitational system. 

\section{Symmetry Classification} \label{section4}

In this section, we present the point symmetries admitted by the parametrization invariant Lagrangian \eqref{Lag}. Leaving aside the infinite-dimensional symmetry $\xi_{\infty}$ represented by \eqref{Xinf} (which every Lagrangian of this form admits) a point symmetry generator for $L$ has the general form
\begin{equation}\label{pointgen}
  \xi = \eta^0(N,a,\phi,X) \frac{\partial }{\partial N} + \eta^1(N,a,\phi,X) \frac{\partial }{\partial a} + \eta^2(N,a,\phi,X) \frac{\partial }{\partial \phi} + \eta^3(N,a,\phi,X) \frac{\partial }{\partial X} .
\end{equation}
The next step is to utilize the prolongation formula \eqref{prolog} together with \eqref{prologcoef} for the above generator, for which $\chi=0$. A direct application of the symmetry criterion \eqref{infcrit} results in a relation where various polynomial terms of ``velocities'' ($\dot{N}$, $\dot{a}$, $\dot{\phi}$, $\dot{X}$) appear, while the functions of the components of \eqref{pointgen} depend only on the ``positions'' ($N$, $a$, $\phi$, $X$). As a result, and in order to satisfy the symmetry criterion, it must be demanded that all velocity coefficients vanish \cite{Olver,Stephani}. This forms an over-determined system of partial differential equations for the coefficients $\eta$. A non trivial solution of \eqref{infcrit} then yields a symmetry generator for $L$. This is an algorithmic method which we refrain from analyzing here, however in the appendix \ref{App} we include the complete list of the equations which emerge from this process.

In what follows, we present the solutions admitted by the symmetry criterion and compare them with known results in the literature, while highlighting some novel emerging symmetries not previously reported. We first distinguish two general cases depending on the nature of $f_1(\phi)$, i.e. being constant or dynamical. These correspond to two large families of models: the minimally coupled, where $f_1(\phi)$ is a constant, and the nonminimally coupled to gravity models, with $f_1(\phi)$ non-constant.

\subsection{Minimal coupling with GR: $f_1(\phi)=$const.}

Apart from the obvious translational symmetry, $\partial_\phi$, which is present when $f_2=f_2(X)$, we distinguish the case where three symmetry vectors exist simultaneously
\begin{align} \label{sym1}
  \xi_1 & = \frac{\partial}{\partial \phi} \\
  \xi_2 & = N \frac{\partial}{\partial N} + \frac{a}{3} \frac{\partial}{\partial a} + \frac{\nu -1 }{\nu } \phi \frac{\partial}{\partial \phi} -\frac{2 X}{\nu } \frac{\partial}{\partial X} \\
  \xi_3 & = N a^{\frac{3 (1-\nu)}{2 \nu -1 }} \frac{\partial}{\partial N} + \frac{2 \nu -1 }{3} a^{\frac{2-\nu }{2 \nu -1}} \frac{\partial}{\partial a} -2 a^{\frac{3 (1-\nu)}{ 2 \nu-1}}X  \frac{\partial}{\partial X},
\end{align}
where the last symmetry, $\xi_3$, appears only when $\nu\neq 1/2$ . These are the symmetries of a theory
\begin{equation} \label{theor1}
  f_1(\phi)=\frac{1}{2\kappa}, \quad f_2(\phi,X) = X^\nu ,
\end{equation}
where we use $\kappa$ to represent the gravitational constant. Notice, that $\xi_2$ also requires $\nu\neq0$ which we assume to be so, otherwise there would be no kinetic term.

Scaling symmetries like $\xi_2$ are usually obtained even through the more restrictive process described in Section \ref{section2}, where $N$ is set to unity before the symmetry calculation, and indeed such a symmetry has been reported previously (see \cite{Capo}). However, the symmetry $\xi_3$ is missed if the lapse is set to unity prior to applying the symmetry criterion to the Lagrangian. The symmetry vector $\xi_3$ is a genuinely new symmetry that has not been reported before in the literature for this class of theories.

The reason why a scaling symmetry like $\xi_2$ can be obtained even from the gauge fixed version of the Lagrangian lies in the form of the infinite dimensional symmetry described by \eqref{Xinf}. When $N$ is set equal to unity, it is eliminated from the Lagrangian as a variable, and the contribution of the  $\frac{\partial}{\partial N}$ component in the generator is removed. However, this missing contribution is not essential if it does not contain variables other than $N$. To see this, consider the vector $\xi_{\infty}+\xi_{2}$ for the particular function $\chi(t)=t$, the resulting vector is
\begin{equation}
  \xi_{\infty}+\xi_{2} \overset{\chi=t}{=} t\frac{\partial}{\partial t} + \frac{a}{3} \frac{\partial}{\partial a} + \frac{\nu -1 }{\nu } \phi \frac{\partial}{\partial \phi} -\frac{2 X}{\nu } \frac{\partial}{\partial X},
\end{equation} 
which is exactly the symmetry obtained for the gauge fixed Lagrangian as a scaling symmetry. On the other hand, $\xi_3$ has a $\frac{\partial}{\partial N}$ component where the variable $a$ appears, and its action cannot be reproduced with a (local) $\chi(t)$ function. Thus, in the search for point symmetries, this symmetry is lost if gauge fixing is performed at the level of the Lagrangian.

Let us now move to a different class of theories, characterized by
\begin{equation} \label{pretheor2}
  f_1(\phi)=\frac{1}{2\kappa}, \quad f_2(\phi,X) = e^{-2h(\phi)} f\left(e^{2 h(\phi )} h'(\phi )^2 X \right) ,
\end{equation} 
where $h$, $f$ are arbitrary functions of their arguments. The prime throughout this work is used to denote differentiation with respect to the argument, in this case $\phi$. The theories of the form given above possess only one symmetry vector:
\begin{equation} \label{presym2}
  \xi_4 = N \frac{\partial}{\partial N} + \frac{a}{3} \frac{\partial}{\partial a} + \frac{1}{s'(\phi)}\frac{\partial}{\partial \phi}  - 2 X \left(1 +\frac{s''(\phi )}{s'(\phi )^2}\right) \frac{\partial}{\partial X} .
\end{equation}
We notice, however, that the arbitrariness in $s(\phi)$ does not really represent different theories, as it is subject to internal transformations of the scalar field. Indeed, by performing the transformation $\phi \rightarrow \exp(s(\phi))$, which induces the change $e^{2 s(\phi )} s'(\phi )^2 X\rightarrow X$, the theory together with its symmetry vector become
\begin{equation} \label{theor2}
  f_1(\phi)=\frac{1}{2\kappa}, \quad f_2(\phi,X) = \frac{1}{\phi^2} f\left(X \right) 
\end{equation}
and
\begin{equation} \label{sym2}
  \xi_4 = N \frac{\partial}{\partial N} + \frac{a}{3} \frac{\partial}{\partial a} + \phi \frac{\partial}{\partial \phi}   .
\end{equation}
It is understood that the $\phi$ and $X$ appearing in \eqref{theor2} and \eqref{sym2} are not the same as those of \eqref{pretheor2} and \eqref{presym2}, as in the former case they are the transformed quantities; we maintain the same symbols just to avoid overburdening the notation. Symmetry \eqref{sym2}, being a simple scaling, has been encountered before in the literature in the context of the class of theories \eqref{theor2} (see \cite{Capo}). We should mention that there is a minor typo in the presentation of this theory in reference \cite{Capo}, where $\phi$ appears instead of $\phi^2$. In addition, the rest of the theories mentioned there for $f_1(\phi)=$const. as distinct cases are in reality equivalent to \eqref{theor2} under re-parametrizations of the scalar field. Just as in our setup, \eqref{pretheor2} is the same theory as \eqref{theor2}, for the same function $f$. Last but not least, it is easy to observe that when $f$ is a linear function, one of the three symmetries for the quintessence/phantom model with exponential potential is recovered \cite{anexp}.

\subsection{Non-minimal coupling with GR: $f_1(\phi)\neq$const.}

In the case where we have a non-constant coupling of the scalar field with the Ricci scalar in the action, we obtain a symmetry for a theory of the form
\begin{equation} \label{pretheor3}
  f_1=f_1(\phi), \quad\quad f_2(\phi,X) = -\frac{3 f_1'(\phi )^2}{f_1(\phi )} X  +f\left(\frac{\left(f_1(\phi)h'(\phi )-2 h(\phi ) f_1'(\phi )\right)^2}{f_1(\phi) h (\phi )^3}X \right),
\end{equation}
where $f_1(\phi)h'(\phi )-2 h(\phi ) f_1'(\phi )\neq 0$, as otherwise the theory reduces to pure quintessence. The symmetry vector is
\begin{equation} \label{presym3}
  \begin{split}
  \xi_5 = & \frac{h f_1'-f_1 h'}{2 h f_1'-f_1 h'} N \frac{\partial}{\partial N}  -\frac{h f_1'+f_1 h'}{6 h f_1'-3 f_1 h'} a \frac{\partial}{\partial a} \\
  & + \frac{2 f_1 h}{2 h f_1'-f_1 h'}\frac{\partial}{\partial \phi}  +\frac{2 \left(\left(3 f_1' h'+2 f_1 h''\right)f_1 h +2  \left(f_1'^2-2 f_1 f_1''\right)h^2-3 f_1^2 h'^2\right)}{\left(f_1 h'-2 h f_1'\right)^2}X \frac{\partial}{\partial X} ,
  \end{split}
\end{equation}
with $f_1$, $h$ and $f$ being arbitrary functions of their arguments. As before,  the above expressions can be significantly simplified by making use of the re-parametrization freedom of the scalar field. Notice however that in this case we have two arbitrary functions of $\phi$: the coupling term with gravity $f_1(\phi)$, and $h(\phi)$. We can perform a transformation in $\phi$ that sets one of the functions to a specific expression in terms $\phi$, but not both of them. Consequently, we obtain here a truly infinite set of theories admitting a symmetry generator. This is a new result in the context of this family of theories, as we will demonstrate however in the subsequent section it is conformally related to the already known symmetry, $\xi_4$, described before.

We choose to transform $\phi$ to fix the function $f_1$. To this end, we redefine the scalar field as $\phi \rightarrow \sqrt{f_1(\phi )}$, which is basically equivalent to setting $f_1=\phi^2$ in the above expressions. The theory and the symmetry vector simplify to:
\begin{equation} \label{theor3}
  f_1=\phi^2, \quad\quad f_2(\phi,X) = -12 X +h(\phi ) f\left(\frac{\left(\phi  h'(\phi )-4 h(\phi )\right)^2}{h(\phi )^3}X \right). 
\end{equation}
The symmetry vector is now given by
\begin{equation} \label{sym3}
  \begin{split}
  \xi_5 = & \frac{2 h-\phi h'}{4 h-\phi h'} N \frac{\partial}{\partial N}  -\frac{\phi h'+2 h}{12 h-3 \phi h'} a \frac{\partial}{\partial a} \\
  & + \frac{2 \phi h}{4 h-\phi h'} \frac{\partial}{\partial \phi}  + \frac{2 \phi \left(2 h \left(\phi h''+3 h'\right)-3 \phi h'^2\right)}{\left(\phi h'-4 h\right)^2} X \frac{\partial}{\partial X} .
  \end{split}
\end{equation}
As we observe, only one arbitrary function of $\phi$ remains, namely the $h(\phi)$, which we can use to distinguish different theories. Of course there is also the arbitrariness of the function $f$ appearing in \eqref{theor3}. We notice that in order for the symmetry to exist, we need to have $4 h-\phi h'\neq 0$, which excludes the case $h(\phi) = \alpha \phi^4$. When $h(\phi) \propto \phi^4$, the argument of $f$ vanishes and the theory reduces to $f_1 = \phi^2$ and $f_2 = -12X + \lambda_0 \phi^4$, which corresponds to the exceptional Weyl invariant scalar-tensor theory with a quartic potential.

As an illustrative example, let us consider the case where $h(\phi)$ is given in terms of a power-law, $h(\phi) = \phi^\mu$. The theory and the symmetry vector become:
\begin{equation} \label{theor3ex}
  f_1=\phi^2, \quad\quad f_2(\phi,X) = -12 X + \phi^\mu f\left(\frac{(\mu -4)^2}{ \phi ^{\mu }} X \right), \quad \mu\neq 4
\end{equation}
and
\begin{equation} \label{sym3ex}
  \xi_{5a} =  (\mu -2) N \frac{\partial}{\partial N}  + \frac{ \mu +2}{3}  a \frac{\partial}{\partial a} - 2 \phi \frac{\partial}{\partial \phi}  - 2\mu X \frac{\partial}{\partial X} ,
\end{equation}
where we have multiplied the symmetry vector by the constant $\mu-4$ to simplify the final expression. This is the case where the symmetry vector reduces to a simple scaling symmetry. In the case where $f$ is a smooth function, by performing an expansion around $X=0$ we can write
\begin{equation} \label{sym3bex}
  f_2(\phi,X) = \sum_{n=0}^{+\infty} a_n \phi^{(1-n)\mu} X^n = a_0 \phi^\mu + a_1 X + a_2 \phi^{-\mu} X^2+..., \quad \mu\neq 4,
\end{equation}
where we absorbed the linear term $-12X$ of \eqref{theor3ex} into the coefficient $a_1 X$. This theory corresponds to a power-law potential $V(\phi)\sim \phi^\mu$ in the Brans-Dicke model when $f_2$ is a function linear  in $X$ \cite{ndsc}.

As a second example, consider the case $h=e^{\mu\phi}$ leading to
\begin{equation} \label{theor3ex2}
  f_1=\phi^2, \quad\quad f_2(\phi,X) = -12 X + e^{\mu \phi} f\left(e^{-\mu  \phi } (\mu  \phi -4)^2 X  \right),
\end{equation}
and
\begin{equation} \label{sym3ex2}
  \xi_{5b} = \frac{\mu  \phi -2}{\mu  \phi -4} N \frac{\partial}{\partial N}  + \frac{\mu  \phi +2}{3 (\mu  \phi -4)} a \frac{\partial}{\partial a} + \frac{2 \phi }{4-\mu  \phi } \frac{\partial}{\partial \phi}  -\frac{2 \mu  \phi  (\mu  \phi -6)}{(\mu  \phi -4)^2} X \frac{\partial}{\partial X} ,
\end{equation}
which is not a scaling symmetry. Performing the same expansion for $f$ in $f_2(\phi,X)$, we obtain 
\begin{equation} \label{sym3bex2}
  f_2(\phi,X) = -12 X+ \sum_{n=0}^{+\infty} a_n (\mu  \phi -4)^{2n} e^{\mu(1-n) \phi} X^n = a_0 e^{\mu \phi} +  a_1 \left[(\mu  \phi -4)^{2}-\frac{12}{a_1}\right]  X + a_2 (\mu  \phi -4)^{4} e^{- \mu \phi} X^2+... \,.
\end{equation}
Notice that if we perform a transformation of the scalar field that ``absorbs'' $[(\mu  \phi -4)^{2}-\frac{12}{a_1}]$ into the kinetic term $X$, the theory will not be the same as the one represented by \eqref{sym3bex}. Different choices of the function $h(\phi)$ represent, in general, distinct theories.

Finally, we distinguish a different case of symmetry, which occurs for a theory of the form
\begin{equation} \label{theor4}
  f_1=\phi^2, \quad\quad f_2(\phi,X) = -12 X +h(\phi ) X^\lambda + V_0 \, \phi^4, \quad \lambda \neq 0,1, 
\end{equation}
where $V_0$ is a constant and $h(\phi)$ an arbitrary non-zero function. The symmetry generator is given by
\begin{equation} \label{sym4}
  \xi_6 = \frac{\phi^{\frac{2}{\lambda }-2}}{h(\phi)^{\frac{1}{2 \lambda }}} N \frac{\partial}{\partial N}  + \frac{\phi^{\frac{2}{\lambda }-2}}{h(\phi)^{\frac{1}{2 \lambda }}} a \frac{\partial}{\partial a} - \frac{\phi^{\frac{2}{\lambda }-1}}{h(\phi)^{\frac{1}{2 \lambda }}} \frac{\partial}{\partial \phi}  + \frac{\phi^{\frac{2}{\lambda }-2}}{\lambda\, h(\phi)^{\frac{1}{2 \lambda }-1}}  \left(\phi h'(\phi)-4 h(\phi)\right) X \frac{\partial}{\partial X} .
\end{equation}
In a certain sense, this case yields the necessary conditions for the potential  $V(\phi) \sim \phi^4$, which was excluded from \eqref{sym3bex}, to have a symmetry. In the special case where the potential is eliminated, i.e. when setting $V_0=0$, the action admits two extra symmetries
\begin{equation} \label{sym5}
  \begin{split}
  \xi_7  =  & \left[1 + \frac{\lambda-1}{\lambda} \frac{\phi^{\frac{2}{\lambda }-2} k(\phi )}{ h(\phi )^{\frac{1}{2 \lambda }}} \right]N \frac{\partial}{\partial N} + \left[ \frac{1}{3} + \frac{\lambda-1}{\lambda} \frac{\phi^{\frac{2}{\lambda }-2} k(\phi )}{ h(\phi )^{\frac{1}{2 \lambda }}} \right]a \frac{\partial}{\partial a} -\frac{\lambda -1}{\lambda } \frac{\phi ^{\frac{2}{\lambda }-1} k(\phi )}{ h(\phi )^{\frac{1}{2 \lambda }}} \frac{\partial}{\partial \phi} \\
  & + \frac{1}{\lambda^2} \left[ (\lambda -1) \phi ^{\frac{2}{\lambda }-2} k(\phi ) h(\phi )^{-\frac{1}{2 \lambda }-1} \left(\phi  h'(\phi )-4 h(\phi )\right)-2 \lambda \right]  X \frac{\partial}{\partial X}
  \end{split}
\end{equation}
and
\begin{equation}
  \xi_8 =(a \phi )^{\frac{3 (1-\lambda )}{2 \lambda -1}} N \frac{\partial}{\partial N} + \frac{1}{3}  (2 \lambda -1) (a \phi )^{\frac{3 (1-\lambda )}{2 \lambda -1}} a  \frac{\partial}{\partial a} -2 (a \phi )^{\frac{3 (1-\lambda )}{2 \lambda -1}} X \frac{\partial}{\partial X},
\end{equation}
where the function $k(\phi)$ appearing in the coefficients of $\xi_7$ is equal to
\begin{equation}
  k(\phi) = -\int \phi ^{1-\frac{2}{\lambda }} h(\phi )^{\frac{1}{2 \lambda }} \, d\phi .
\end{equation}
In the following section we establish the connection of these last symmetries to the known $\xi_2$ and the new $\xi_3$ symmetries respectively. In Table \ref{tab:symmetries}, we summarize our results, including the various classes of theories admitting point symmetries.

\begin{table}[ht]
\centering
\begin{tabular}{|c|c|c|}
  \hline
  % after \\: \hline or \cline{col1-col2} \cline{col3-col4} ...
  Theory & Symmetry vectors & Conditions\\ \hline\hline
  $f_{1}=1/(2\kappa), \quad f_2= f_2(X)$ & $\partial_\phi$ & - \\ \hline
  $f_{1}=1/(2\kappa), \quad f_2= X^\nu$ & $\xi_1, \xi_2, \xi_3$ & $\nu\neq 0, 1/2$ \\ \hline
  $f_{1}=1/(2\kappa), \quad f_2= f(X)/\phi^2$ &  $\xi_4$ (scaling) & - \\ \hline
  $f_{1}=\phi^2, \quad f_2= -12 X +h(\phi ) f\left( X \left(\phi  h'(\phi )-4 h(\phi )\right)^2/h(\phi )^3 \right)$ & $\xi_5$ & $h(\phi) \neq \alpha \phi^4$  \\ \hline
  $f_{1}=\phi^2, \quad f_2=  -12 X +h(\phi ) X^\lambda + V_0 \, \phi^4$ & $\xi_6$ & $\lambda \neq 0,1$, $h(\phi)\neq0$ \\ \hline
  $f_{1}=\phi^2, \quad f_2=  -12 X +h(\phi ) X^\lambda$ & $\xi_6, \xi_7, \xi_8$ & $\lambda \neq 0,1$, $h(\phi)\neq0$ \\ \hline
\end{tabular} %}
\caption{The distinct cases of $k$-essence theories admitting point symmetry generators.}
\label{tab:symmetries}
\end{table}

\subsection{Conformal mappings}

In this section we explore the correspondence among the symmetries of theories with $f(\phi)\neq$const. with those where $f(\phi)$ is constant. Under a conformal transformation of the metric
\begin{equation}
   \tilde{g}_{\mu\nu} = \Omega^2 g_{\mu\nu}
\end{equation}
The Ricci scalar transforms as \cite{Wald}
\begin{equation}
  \tilde{R} = \Omega^{-2} \left[ R - 6 \nabla_\mu\nabla^\mu \ln\Omega - 6 \nabla_\mu (\ln\Omega) \nabla^\mu (\ln\Omega)  \right].
\end{equation}
Taking into account that $\sqrt{-\tilde{g}}=\Omega^4 \sqrt{-g}$ and leaving out a term corresponding to a total derivative, the usual Einstein-Hilbert lagrangian density maps to
\begin{equation}
    \sqrt{-\tilde{g}}\tilde{R} \mapsto \sqrt{-g} \left( \Omega^2 R + 6 \nabla_\mu \Omega \nabla^\mu \Omega \right)
\end{equation}
By choosing $\Omega= f_1(\phi)$ theories with $f_1(\phi)=$const. can be mapped to the theories with a non constant coupling.

It is important to note that conformally related theories are not physically equivalent, as conformal transformations in general affect the behavior of the gravitational field (for example, singular spacetimes can become regular under a conformal transformation). Nevertheless it is a mathematically valid transformation which consequently can map symmetries of one physical theory to the other. 

Let us take for example, the vector $\xi_5$ of \eqref{sym3}. The transformation $g_{\mu\nu}= \frac{1}{f_1(\phi)} \tilde{g}_{\mu\nu}$ corresponds to
\begin{equation}\label{cftr1}
   N = \frac{\tilde{N}}{\sqrt{f_1(\phi)}}, \quad a = \frac{\tilde{a}}{\sqrt{f_1(\phi)}}, \quad X= f_1(\phi) \tilde{X} ,
\end{equation}
where for the last relation we took into account the fact that 
\begin{equation}
  X= - \frac{1}{2}g^{\mu\nu} \partial_\mu \phi \partial_\nu \phi = - \frac{f_1(\phi)}{2}\tilde{g}^{\mu\nu} \partial_\mu \phi \partial_\nu \phi= f_1(\phi) \tilde{X} .
\end{equation}
By calculating the Jacobian $J^K_L = \frac{\partial \tilde{Q}^K}{\partial Q^L}$ and applying the above transformation on the vector $\xi_5= \xi_5^K \frac{\partial}{\partial Q^K}$, we obtain the transformed vector $\tilde{\xi}_5 = \tilde{\xi}_5^K \frac{\partial}{\partial \tilde{Q}^K}$ with components $\tilde{\xi}_5^K = \frac{\partial \tilde{Q}^K}{\partial Q^L} \xi_5^L$ resulting in
\begin{equation}
  \tilde{\xi}_5 = \tilde{N} \frac{\partial}{\partial \tilde{N}} + \frac{\tilde{a}}{3} \frac{\partial}{\partial \tilde{a}} + \frac{1}{s'(\phi)}\frac{\partial}{\partial \phi}  - 2 \tilde{X} \left(1 +\frac{s''(\phi )}{s'(\phi )^2}\right) \frac{\partial}{\partial \tilde{X}} ,
\end{equation}
where $f_1(\phi)$ and $h(\phi)$ of \eqref{presym3} are related to the $s(\phi)$ through
\begin{equation} \label{relation1}
   s'(\phi ) = \frac{f_1'(\phi )}{f_1(\phi ) } -\frac{h'(\phi )}{2 h(\phi )} . 
\end{equation} 
The vector $\tilde{\xi}_5$ above is exactly the symmetry vector $\xi_4$ of \eqref{presym2} of the minimally coupled theory. Equation \eqref{relation1} establishes the relation between the fundamental functions of the theory needed in order to realize the correspondence.

Let us now verify what happens with the other symmetry associated with a nonminimally coupled term in the theory, namely the vector $\xi_6$ of \eqref{sym4}. By applying the exact same transformation we see in \eqref{cftr1} with $f_1(\phi)=\phi^2$ we obtain a transformed vector that reads
\begin{equation}
  \tilde{\xi}_6 = \frac{\phi ^{\frac{2}{\lambda }-2} }{\lambda h(\phi )^{\frac{1}{2 \lambda }+1}} \left(\phi  h'(\phi )+2 (\lambda -2) h(\phi )\right) \tilde{X} \frac{\partial}{\partial \tilde{X}}  - \frac{\phi ^{\frac{2}{\lambda }-1}}{h(\phi )^{\frac{1}{2 \lambda }} } \frac{\partial}{\partial \phi} .
\end{equation}
This vector represents a pure transformation in the space of the scalar field, if we successively enforce the change 
\begin{equation}
  \phi\mapsto \tilde{\phi} =-\int \phi ^{1-\frac{2}{\lambda }} h(\phi )^{\frac{1}{2 \lambda }} d\phi,
\end{equation}
and of course taking into account the effected transformation in $X$ the vector reduces to
\begin{equation}
  \tilde{\xi}_6 = \frac{\partial}{\partial \tilde{\phi}},
\end{equation}
which is the translation symmetry seen in \eqref{sym1}. It is particularly interesting that such a simple translational symmetry in nonminimally coupled theory without $\phi$ dependence can manifest itself as something quite elaborate in a conformally related theory yielding a nontrivial conserved quantity. 

Under the exact same transformation it can be seen that $\xi_7$ is associated with $\xi_2$, while $\xi_8$ is the conformally related version of $\xi_3$. Hence we obtain a complete correspondence in the two frames. In what follows we explore some particular solutions that can be obtain with the use of the presented symmetries.

\section{Particular solutions} \label{section5}

In this section we make use of the conservation laws implied by the previously reported symmetries to derive some exact solutions.

\subsection{The scaling symmetry $\xi_4$}

Let us first concentrate on the theory \eqref{theor2}, which admits the symmetry \eqref{sym2}. The phase-space dynamics of a particular theory admitting this symmetry were recently studied in \cite{ka2}. The resulting conserved charge is
\begin{equation} \label{examplesolI1}
  I_{sc} = \frac{3 a^2}{\kappa  N \phi} \left(\kappa  a \dot{\phi} f_2'(X)-2 \phi \dot{a}\right),
\end{equation}
where the prime here denotes derivation with respect to the argument $X$. It can be easily verified that this quantity is conserved since, upon substituting the accelerations, we obtain
\begin{equation}
  \frac{dI_{sc}}{dt} = - 3 N \frac{\partial L}{\partial N}.
\end{equation}
On the right-hand side we have the constraint, which vanishes on shell, thus, $\frac{dI_{sc}}{dt} =0$. 

In the particular case where $I_{sc}=0$, the expression \eqref{examplesolI1} and the definition of the kinetic term \eqref{defX} imply
\begin{align} \label{examplesol1N}
  N & = \frac{\dot{\phi}}{\sqrt{2} \sqrt{X}}, \\ \label{examplesol1a}
  \frac{\dot{a}}{a} &= \kappa\frac{  \dot{\phi}}{2 \phi} f_2'(X) .
\end{align}
Using these relations in the equations of motion, it can be easily seen that the latter are satisfied if
\begin{equation} \label{example1alg}
  X f_2'(X) \left(3 \kappa  f_2'(X) -4\right)+2 f_2(X) = 0.
\end{equation}
This implies that either the theory is fixed to be of the form
\begin{equation}
  f_2(X) = \frac{2}{3} C_1 \sqrt{X}-\frac{C_1^2 \kappa}{6} ,
\end{equation}
with $C_1$ being a constant of integration, or the theory is not specified, i.e. function $f_2(X)$ is not assigned to a particular form, with $X$ being a constant satisfying the algebraic equation \eqref{example1alg}. Of course, in order to have a physically valid solution the corresponding equation needs to have real nonnegative roots, as required for a homogeneous canonical scalar field, and $f_2(X)$ to be a real valued function appropriate to describe a physical system. We need to mention, that apart from simple polynomial choices for $f_2(X)$, equation \eqref{example1alg} might not be analytically solvable. In this respect it may be easier to think of the latter as a defining equation for the parameters of the theory, and not of $X$. A trivial example is if we consider a theory of the form $f_2(X) = f_3(X) + \Lambda$. Then, \eqref{example1alg} simply reduces for a defining equation for the cosmological constant
\begin{equation}
  \Lambda = \frac{1}{2} X f_3'(X) \left(4-3 \kappa  f_3'(X)\right)-f_3(X).
\end{equation}
Similarly, it may be more relevant to impose functions family of functions $f_2(X)$ characterized couplings, whose range of values can be obtained from equation \eqref{example1alg}.

In this latter $X=$const. case, we deduce from \eqref{examplesol1N}, that $\phi$ effectively becomes the time variable, since $N dt \propto \dot{\phi} dt = d\phi$. This implies that there exists a solution with $X=$const. $N=$const. and $\phi=t$. Truly, if we take $X$ as a constant it is easy to derive that the set
\begin{align}
   N& =1, \quad\quad a = a_0 \, t^{\frac{\kappa}{2}f'(X)} ,\\
   \phi & = \left(2 X\right)^{1/2} t,  \quad\quad   X = \text{constant}
\end{align} 
satisfies the field equations, with $X$ bound by the algebraic constraint \eqref{example1alg}. Thus, we see that a power-law type of solution for the scale factor is always admitted by this family of theories, with the specific exponent being decided by the functional dependence of $f_2$ on $X$.

\subsection{The case of $\xi_5$: A common solution for an infinite group of theories.} 

The symmetry $\xi_5$ given by \eqref{sym3} exists for the family of theories \eqref{theor3}, where, as we previously mentioned, $h(\phi)$ can be an arbitrary function of $\phi$, save for the case that trivializes the denominators. The theory is conformally related to that of the previous case and we work in a similar manner starting from the conserved charge, where in this frame is
\begin{equation}
  I = -\frac{3 a^2 \phi }{N h (\phi)} \left[2 h (\phi ) \left(\phi \, \dot{a} + a \dot{\phi} \left(1-2 \frac{df(u)}{du} \right)\right)+a \phi  \dot{\phi} h '(\phi) \frac{df(u)}{du} \right],
\end{equation}
where
\begin{equation} \label{defu}
  u = \frac{X \left(\phi\, h'(\phi)-4 h(\phi) \right)^2}{h(\phi)^3} .
\end{equation}
The total time derivative of $I$, after substitution of all accelerations, leads to a relation of the form
\begin{equation}
   \frac{dI}{dt} = N A(\phi) \frac{\partial L}{\partial N} + B(N,a,\phi,X) \left( X - \frac{\dot{\phi}^2}{2 N^2} \right) ,
\end{equation}
which vanishes on shell. The first term on the right-hand side is the constraint $\frac{\partial L}{\partial N}=0$ and the second is the definition of the kinetic term \eqref{defX}, which emerges here as a constraint due to the Lagrange multiplier. 

Once more, for the particular case $I=0$, we are led to an exact solution. In order to simplify the expressions let us re-parametrize $h(\phi)$ and $N(t)$ as 
\begin{align}
   h(\phi)& = \frac{2 \phi^4}{S(\phi)^2}, \\
   N(t)& = \pm \frac{S'(\phi)}{\sqrt{u(t)} \phi}\dot{\phi} ,
\end{align}
introducing the new function $S(\phi)$ and using the $u(t)$ of \eqref{defu}. Then, the $I=0$ equation conveniently reduces to 
\begin{equation}
  \frac{\dot{a}}{a} = \left(\frac{S'(\phi)}{S(\phi)} \frac{df(u)}{du}  -\frac{1}{\phi}\right) \dot{\phi} .
\end{equation}
Substitution of the above into the equations of motion reveals the condition
\begin{equation} \label{alg2eq}
  u \frac{df(u)}{du} \left(3 \frac{df(u)}{du}-2\right)+f(u) =0 .
\end{equation}
As in the previous case, this can either be satisfied for a theory
\begin{equation}
  f(u) = \frac{2 C_1}{3} \sqrt{u} -\frac{C_1^2}{3},
\end{equation}
or with $u$ being a constant satisfying the algebraic equation \eqref{alg2eq} for some given $f(u)$ theory. Similarly to the case of \eqref{example1alg}, it may be more relevant use equation \eqref{alg2eq} as one that dictates the values or relations among the coupling constants in the theory, instead of specifying a specific $u$ value.

In the case where $u=$const., $\phi$ becomes once more the effective time variable $t$, and we are able to write the solution
\begin{align}
   & N(t) = \pm \frac{\dot{S}}{t \sqrt{u}}, \\
   & a(t)= a_0 \frac{ S(t)^{f'(u)}}{t}, \\
   & \phi(t)=t , \\
   & X(t) = \frac{u\,  t^2}{2 \dot{S}^2}
\end{align}
where $a_0$ is a constant. In this case, $S(\phi)$ is converted to $S(t)$ and continues to be an arbitrary function, while $u$ has to be a root of the algebraic equation \eqref{alg2eq}. Thus, the above solution can be applied to infinitely many theories as neither $f(u)$, nor $S(\phi)$ are fixed. 

\subsubsection{A particular example}

As a simple example let us see what happens if we choose
\begin{equation} \label{choiceS}
  S(\phi) = \frac{\sqrt{2}}{\mu-4} \phi^{\frac{4-\mu }{2}}.
\end{equation}
This selection corresponds to a theory of the form
\begin{equation} \label{example2}
  f_2(\phi,X) = -12 X + (\mu -4)^2 \phi^{\mu } f\left(\phi^{-\mu } X \right),
\end{equation}
with solution
\begin{align} \label{solex2gen}
  & N(t) = \mp \frac{t^{-\frac{\mu }{2}}}{\sqrt{2} \sqrt{u}}, \\
  & a(t)= a_0 \, t^{\frac{1}{2} (4-\mu) f'(u )-1} , \\
  & \phi=t ,\\
  & X=u \, t^\mu
\end{align}
where $u$, of course, must be a constant root of Eq. \eqref{alg2eq} for some chosen theory $f(u)$. For the choice \eqref{choiceS} and the family of theories \eqref{example2}, this is also a scaling solution, since upon transforming to the cosmic time gauge $N(t) dt = d\tau$ the solution reads (if $\mu\neq 2$):
\begin{align}
  & N(\tau)=1, \\
  & a(\tau)= \tilde{a}_0 \, \tau^{\frac{(\mu -4) f'(u )+2}{\mu -2}} ,\\
  & \phi(\tau) = 2^{\frac{1}{\mu -2}} |\mu -2|^{-\frac{2}{\mu -2}} u ^{-\frac{1}{\mu -2}} \tau^{-\frac{2}{\mu -2}}, \\
  & X(\tau)=2^{\frac{\mu }{\mu -2}} |\mu -2|^{-\frac{2 \mu }{\mu -2}} u ^{1-\frac{\mu }{\mu -2}} \tau^{-\frac{2 \mu }{\mu -2}} ,
\end{align}
where $\tilde{a}_0$ is a new constant related by a scaling to $a_0$ and the absolute value $|\mu -2|$ is obtained by appropriately choosing the plus or minus sign of $N(t)$ in \eqref{solex2gen} so that the transformation $t\rightarrow \tau$ remains real. It is understood that, as previously discussed, $u$ is constrained to be a constant root of \eqref{alg2eq}. 

In the case where $\mu=2$, the transformation from $t$ to $\tau$, in order to express the solution in the cosmic gauge, follows a different branch, and we obtain a de Sitter solution:
\begin{align}
  & N(\tau)=1, \\
  & a(\tau)= \tilde{a}_0 \, e^{\sqrt{2u} \left(f'(u)-1\right) \tau } ,\\
  & \phi(\tau) =e^{\sqrt{2} \sqrt{u} \tau}, \\
  & X(\tau)=u\,  e^{2 \sqrt{2} \sqrt{u } \tau} .
\end{align}

It is interesting to note that the general form of the solution holds for any choice of the function $f$, with the latter affecting, through the roots of Eq. \eqref{alg2eq}, the numerical values of the exponents in the power-law.

\section{Conclusions} \label{section6}

We have performed a detailed classification of the $k$-essence cosmological models based on the admitted variational symmetries. The kinetic term $X$ was treated as an independent degree of freedom through a Lagrange multiplier, while the lapse function $N$ was retained as a dynamical variable in the minisuperspace. As a result, the symmetry analysis was performed on the parametrization invariant cosmological system. By allowing a free lapse function $N$, we were able to identify novel families of models possessing nontrivial symmetries that had not been previously reported in the literature. Although we restricted our analysis to pure $k$-essence theories, that is, $f_{2,XX}\neq0$, our results consistently recover earlier findings for quintessence and scalar-tensor theories. 

The $k$-essence models with nontrivial symmetries were classified into two broad families: the minimally coupled and the nonminimally coupled to gravity theories. For the minimally coupled to gravity models, with $f_1=1/(2\kappa)$ and $f_2=X^\nu$ the field equations possess three variational symmetries. The first two correspond to a shift of the scalar field and to a scaling symmetry. However, the third symmetry vector depends on the scale factor and it has not been presented before in the literature. Moreover, we found a family of models  admitting a scaling symmetry that includes the quintessence scalar field theory with and the exponential potential. For the second family of nonmininally coupled to gravity models, we derive an infinite set of theories that admit a symmetry vector, which reduces to a scaling symmetry for specific choices of the free functions. Non-scaling symmetries are those which are lost when a non-parametrization invariant approach is followed in the symmetry analysis. Finally, we applied the Noetherian conservation laws to derive exact solutions to the cosmological field equations.

We additionally explored the conformal transformations that connect different families of theories. While the conformal mappings utilized in this work depend solely on the scalar field, extending these transformations to include dependencies on the kinetic term $X$, or even to include disformal transformations \cite{disf}, presents an intriguing avenue for future research. Such generalized transformations could map the identified $k$-essence symmetries to entirely different families of gravitational theories, such as Horndeski or beyond-Horndeski models. Investigating how these broader phase-space mappings preserve, alter, or generate new symmetry generators and their corresponding Noetherian charges could provide deeper insights into the underlying structural connections between distinct extended theories of gravity and their cosmological dynamics.

The symmetry analysis presented here provides a geometric selection rule for the otherwise arbitrary functions of $k$-essence theories. In a future study, we plan to extend this analysis by using the resulting conservation laws to identify quantum observables in the canonical quantization of such theories along the lines presented before in \cite{ka1}. Furthermore, the physical properties of the derived models, as well as the physical interpretation of the Noetherian conservation laws, will be discussed elsewhere.

\appendix

\section{The symmetry criterion equations} \label{App}

As mentioned in the main text, we need to find the coefficients of the symmetry vectors satisfying \eqref{infcrit}. For a point symmetry of the form \eqref{pointgen}, and for Lagrangian \eqref{Lag}, we obtain an equation that contains polynomial terms of the first derivatives $\dot{N}$, $\dot{a}$, $\dot{\phi}$ and $\dot{X}$. Given that the coefficients of the symmetry vector \eqref{pointgen} do not depend on these derivatives, this implies that the factors multiplying the various polynomial terms must be separately zero. This defines an overdetermined system of equations. In what follows we present the derived equation out of each polynomial term. We consider a trivial gauge function $F=$const. in \eqref{infcrit}.

The equations out of the multiplying factors of polynomial terms  are (to simplify the symbolism we write the indices indicating the vector components below, i.e. $\eta_I$ instead of $\eta^I$, with $I=0,\ldots,3$):
\begin{itemize}
  \item term $\dot{a}^2$
     \begin{equation}
       a N f_1'(\phi) \left(a \, \eta_{2,a}+\eta_2\right)+f_1(\phi) \left(N \left(2 a\, \eta_{1,a}+\eta_1\right)-a\, \eta_0\right) =0,
     \end{equation}
  \item term  $\dot{\phi}^2$:
     \begin{equation}
        N \left(-12 f_1'(\phi) \eta_{1,\phi}+3 f_{2,X}\, \eta_1+2 a f_{2,X} \, \eta_{2,\phi}+a f_{2,X\phi} \, \eta_2 + a f_{2,XX} \, \eta_3\right)-a f_{2,X}\,  \eta_0 =0,
     \end{equation}
  \item term  $\dot{a} \dot{N}$:
     \begin{equation}
        a f_1'(\phi) \, \eta_{2,N}+2 f_1(\phi) \, \eta_{1,N}=0,
     \end{equation}
  \item term  $\dot{\phi} \, \dot{N}$:
     \begin{equation}
        a f_{2,X} \, \eta_{2,N}-6 f_1'(\phi) \, \eta_{1,N} =0,
     \end{equation}
  \item term  $\dot{a}\, \dot{\phi} $:
     \begin{equation}
        6 a f_1'(\phi) \, \eta_0-N \left[6 a f_1''(\phi) \eta_2 +6 a f_1'(\phi) \left( \eta_{1,a} + \eta_{2,\phi} \right) - a^2 f_{2,X} \, \eta_{2,a} + 12 f_1'(\phi) \eta_1+12 f_1(\phi) \eta_{1,\phi}\right]=0,
     \end{equation}
  \item term   $\dot{a}\, \dot{X} $:
      \begin{equation}
        a f_1'(\phi) \eta_{2,X}+2 f_1(\phi) \eta_{1,X} =0,
      \end{equation}
  \item term  $\dot{\phi}\, \dot{X} $:
     \begin{equation}
        a f_{2,X} \, \eta_{2,X}-6 f_1'(\phi) \eta_{1,X}=0,
     \end{equation}
  \item term $\dot{a}$:
     \begin{equation}
        6 a f_1'(\phi) \eta_{2,t}+12 f_1(\phi) \eta_{1,t}=0,
     \end{equation}
  \item term $\dot{\phi}$:
     \begin{equation}
        a^2 f_{2,X} \, \eta_{2,t}-6 a f_1'(\phi) \eta_{1,t} =0,
     \end{equation}
  \item zeroth order term
     \begin{equation}
       a X f_{2,X} \eta_0 + N \left(3 X f_{2,X} \, \eta_1+a \left(X f_{2,X\phi}-f_{2,\phi}\right) \eta_2+a X f_{2,XX} \, \eta_3\right) - f_2 (a \, \eta_0+3 N \eta_1) =0 .
     \end{equation}
\end{itemize}

\begin{acknowledgments}
AP was partially supported from FONDECYT Grant 1240514. AL acknowledges financial support from Universidad de La Frontera. AP acknowledges the COST Action CA23130 ``Bridging high and low energies in search of quantum gravity (BridgeQG)''.
\end{acknowledgments}

\end{document}